\def\be{\begin{equation}}
\def\ee{\end{equation}}

\documentstyle[aps,epsf,prl]{revtex}

\begin{document}
\twocolumn
[
\draft
\title{Infinite Hierarchy of Exact Equations
in the Bak-Sneppen Model}
\author{Sergei Maslov}
\address{
Department of Physics, Brookhaven National
Laboratory, Upton, New York 11973\\
and Department of Physics, SUNY at Stony Brook, Stony Brook, New York 11794}
\date{\today}
\maketitle

\widetext
\advance\leftskip by 57pt
\advance\rightskip by 57pt

\begin{abstract}
We derive an infinite hierarchy of
exact equations for the Bak-Sneppen model in arbitrary
dimensions. These equations relate different moments of 
temporal duration and spatial size of avalanches. 
We prove that the exponents of the BS model are the same above 
and below the critical point and express the universal amplitude 
ratio of the avalanche spatial size in terms of the critical exponents.
The equations uniquely determine the shape of the scaling function
of the avalanche distribution.
It is suggested that in the BS model 
there is only one independent critical exponent.
\end{abstract}
]
\narrowtext

Recently Bak and Sneppen  \cite{ bsmodel}
introduced a particularly simple toy model of biological evolution
(BS model). The
model correctly reproduces such features of real evolution process 
as punctuated equilibria, power law probability
distributions of lifetimes of species and of the sizes of
extinction events. 
In spite of the simplicity of the rules, the model exhibits 
extremely rich and interesting behavior. 

In the one-dimensional version of the model $L$ numbers 
$f_i$ are arranged on a line. At every time step
the smallest number in the system and its 
nearest neighbors are replaced
with new uncorrelated random
numbers drawn from the uniform distribution between 0 and 1.
The generalization to higher dimensions is straightforward.
In fact there exists a whole class of models where one selects the
site with the extremal (global maximal or minimal)   
value of some variable and then changes this 
variable at this site and its nearest neighbors
according to some stochastic rule. One of the best known
representatives of this class is Invasion
Percolation \cite{ ip}. Such models, referred to as
{\it extremal models} were extensively studied
(for a recent review see \cite{ PMB}).

The interesting feature of the BS model (as well as other
extremal models) is its ability to organize itself into a scale-free
stationary state. The dynamics in this critical
state is given in terms of bursts of activity or {\it avalanches}, 
which form a hierarchical structure \cite{ bsmodel, PMB}
of subavalanches within bigger avalanches.
Here we introduce a ``master'' equation for this 
avalanche hierarchy. It describes the cascade process
of smaller avalanches merging into bigger ones as the 
critical parameter is changed. From this equation 
we derive an {\it infinite series} of {\it exact} equations, relating
different moments of temporal duration $S$
and spatial size $R^d$ of individual avalanches. 

The ``master'' equation connects undercritical and overcritical 
regions of parameters. Given the existence of the scaling, 
we rigorously prove that the exponents
of the BS model are the same above and below the transition. 
From our results it follows that all terms of Taylor series
of the scaling function $f(x)$ for the avalanche distribution are
uniquely and explicitly 
determined by two critical exponents of the model. 
It was suggested that usual restrictions
on the shape of $f(x)$ indirectly relate these two exponents
and, therefore, reduce the
number of independent critical exponents in the BS model
to just one. 

To define these avalanches one records 
the {\it signal} of the model, i.e. the value of the global minimal
number $f_{min}(s)$ as a function of time $s$.
Then for every value of the auxiliary parameter $f_o$, an
$f_o$-avalanche of size (temporal duration) $S$
is defined as a sequence of $S-1$ successive events
when $f_{min}(s)<f_o$ confined between two events 
when $f_{min}(s)>f_o$. In other words, the events when
$f_{min}(s)>f_o$ divide the time axis into a series of 
avalanches, following one another. 
It is easy to see that an avalanche defined by this rule is nothing
else but a creation-annihilation branching process where sites
with $f_i<f_o$ play the role of particles. The avalanche is 
terminated (and the next one is immediately started) when there are 
no such ``particles'' left in the system.
As in any other creation-annihilation branching processes (such as
directed percolation)
in BS model there exists a critical value $f_c$ of the creation
probability $f_o$, 
for which the creation of particles is just marginally balanced by their
annihilation, and avalanches of all sizes can be realized. 
In the stationary state of the BS model on the infinite lattice, 
$f_{min}(s) \leq f_c$ for every $s$. Therefore, the overcritical ($f_o>f_c$)
region of the branching process parameters is not accessible, since
there are no  events starting or terminating such avalanches. 
However, if the system is artificially prepared 
in the overcritical state with $f_i > f_o$ everywhere,  
one can observe overcritical avalanches. In this case there is
a non-zero probability $P_{\infty}(f_o)$ to start an infinite
avalanche and at the same time the size of finite avalanches
has a finite cutoff.

The events within the same avalanche are
spatially and causally connected. It is easy to understand
that the position of active site at any time step within 
the avalanche is connected to the set of sites
covered (updated at least once) by the avalanche up to this time 
step.
We characterize an avalanche by two principal numbers: 
1) $S$ -- the avalanche size, equal to its temporal duration; 
2) $n_{cov}$ -- the number of covered sites, i.e 
sites that had their random number updated at least once
during the course of this avalanche. In one-dimensional
models the connected nature of the set of covered
sites ensure its compactness and, therefore, 
$n_{cov}$ is equal to the avalanche spatial 
extent $R$.
In higher dimensions (below the upper critical dimension)
it was conjectured in \cite{ PMB} 
that the set of covered sites is a non-fractal
object of the same dimensionality $d$ as the underlying lattice. 
In this case the spatial size $R$ of the avalanche can be {\it
defined} by the relation $n_{cov} = R^d$.
                                                                              
$f_o$-avalanches in the Bak-Sneppen model were shown to be {\it exactly}
equivalent to the realizations of the {\it BS branching process} \cite{ gap}. 
In this process one only  keeps track of the numbers $f_i < f_o$, and 
at each time step activates the smallest one of them. 
The BS branching process is terminated when
there are no numbers $f_i < f_o$ left in the system.
Besides the fact that the BS branching process is a very effective
way to simulate the BS model numerically, 
it has the additional important advantage that
the overcritical region $f_o>f_c$ becomes accessible.

 The quantity of primary interest in the BS model 
is the probability distribution $P(S, f_o)$
of the avalanche sizes $S$ at any given value of the auxiliary parameter
$f_o$. The moments in time, when $f_o<f_{min}(s)<f_o+df_o$ serve
as breaking points for $f_o$-avalanches but not for
$f_o+df_o$-avalanches. Therefore, when 
$f_o$ is raised by an infinitesimal amount $df_o$
some of $f_o$-avalanches {\it merge} together to form bigger
($f_o+df_o$)-avalanches. In the rest of this paper
we study in more detail the properties of this merging
process and the avalanche hierarchy that it induces.

The most important observation about $f_o$-avalanches in the BS model
(as well as in several other extremal models, such as the Sneppen model
\cite{ sneppen} or Invasion Percolation \cite{ ip}) 
is that when an $f_o$-avalanche is
terminated, the numbers $f_i$ on the set of $n_{cov}=R^d$
updated sites are {\it uncorrelated and uniformly
distributed between $f_o$ and $1$}. To prove this statement one
notices that at any single time step all new random numbers are drawn
from the uniform distribution between $0$ and $1$. If such a number
happens to be larger than $f_o$, its exact value 
is not important for the dynamics of this avalanche.

The direct consequence of this observation is that the probability of
an $f_o$-avalanche of spatial size $R^d$
to merge with the subsequent one when the parameter $f_o$ is raised by
$df_o$ is given by $R^d {df_o \over 1-f_o}$. 
(the merging occurs if at least one of the changed
numbers falls in $[f_o, f_o+df_o]$.)
For the following arguments to be true 
it is important that any two subsequent
avalanches are mutually uncorrelated. That is: 
the probability distribution 
of $f_o$-avalanches, starting immediately after 
the termination of an $f_o$-avalanche 
of a given size $S$ is independent of $S$. 
That is true for the BS model since the dynamics within an $f_o$-avalanche
in BS model is completely independent of the particular value
of the numbers $f_i>f_o$ in the background that were left by the
previous avalanches.
This mutual independence may {\it not be the case} 
for other extremal models such as the Sneppen model or
Invasion Percolation. To understand to what extent the results
of this work apply to these other models is the direction of our 
current research \cite{ tbp}.

Now we are in a position to write down the {\it exact} 
``master'' equation describing
how avalanche merging changes $P(S, f_o)$ as $f_o$ is raised.
Let $R^d(S,f_o)$ be the average
number of updated (covered) sites in an $f_o$-avalanche of temporal
size $S$. From our simulations of the BS model \cite{ PMB} we know that 
for $f_o$ close to $f_c$, $R^d(S,f_o)$ scales with $S$ as $S^{d/D}$, where
$D$ is the fractal mass dimension of the avalanche. 
However, for the following arguments 
any form of $R^d(S, f_o)$ will suffice.
The ``master'' equation for $P(S, f_o)$ can be written as
\begin{eqnarray}
(1-f_o){\partial P(S,f_o) \over \partial f_o}=-P(S,f_o)R^d(S,f_o)
\nonumber \\
+\sum_{S_1=1}^{S-1} P(S_1,f_o)R^d(S_1,f_o)P(S-S_1,f_o) \qquad .
\label{e.m1}
\end{eqnarray}
Here the first term describes the loss of avalanches of size $S$
due to the merging with the subsequent one, while the second term
describes the gain in $P(S, f_o)$
due to merging of avalanches of size $S_1$
with avalanches of size $S-S_1$.
It is convenient to change variables from $f_o$ to $g=-\ln(1-f_o)$, so
that $f_o=0$ corresponds to $g=0$, $f_o=1$ corresponds to $g=+\infty$, 
and $dg={df_o \over 1-f_o}$. This change is due to the fact that, although 
traditionally new random numbers are drawn
from the flat distribution ${\cal P}(f_o)=1$, 
the ``natural'' distribution for the BS model
has the probability density ${\cal P}(g)=e^{-g}$.
As usual, the critical properties
of the model are independent of the particular shape
of $\cal P$.
In the rest of the paper
we will use the ``natural'' variable $g$ instead of $f_o$.

To proceed further we make the Laplace transformation of $P(S,g)$: 
$p(\alpha ,g)=\sum_{S=1}^{\infty}P(S,g)e^{-\alpha S}$ 
and of $P(S,g)R^d(S,g)$: $r(\alpha ,g)=
\sum_{S=1}^{\infty}P(S,g)R^d(S,g)e^{-\alpha S}$.
The equation (\ref{e.m1}) can be conveniently written 
in terms of Laplace transforms $p(\alpha ,g)=\sum_{S=1}^{\infty}
P(S,g)e^{-\alpha S}$ and $r(\alpha ,g)=
\sum_{S=1}^{\infty}P(S,g)R^d(S,g)e^{-\alpha S}$ as 
${\partial p(\alpha ,g)/ \partial g}=
-r(\alpha ,g)+p(\alpha ,g)r(\alpha ,g)$, or simply,
\be
{\partial \ln(1-p(\alpha ,g)) \over \partial g}=r(\alpha ,g)
\label{e.m2}
\ee
This exact equation is the central result of this paper.
It has many interesting physical consequences.
When $g<g_c$ all avalanches are finite ($P_{\infty}=0$) and  
normalization requires $p(0,g)=\sum_{S=1}^{\infty}P(S,g)=1$. 
From the general properties 
of the Laplace transform one can write the 
Taylor series for $p(\alpha ,g)$ and $r(\alpha ,g)$ at $\alpha=0$ 
as $p(\alpha ,g)=1-\langle S \rangle _g \alpha  +\langle S^2 \rangle _g
\alpha ^2/2 - \langle S^3 \rangle _g
\alpha ^3/6 + \ldots $ and $r(\alpha ,g)=\langle R^d \rangle _g -
\langle R^d S \rangle _g \alpha  + \langle R^d S^2 \rangle _g
\alpha ^2/2 -  + \ldots $.
Substitution of these expressions into
Eq. (\ref{e.m2}) results in ${\partial \over \partial g}
\ln(\langle S \rangle _g 
\alpha  -\langle S^2 \rangle _g \alpha ^2/2 + \ldots)
=\langle R^d \rangle _g -
\langle R^d S \rangle _g \alpha  + \langle R^d S^2\rangle _g 
\alpha ^2/2 + \ldots $. 
Since the equation (\ref{e.m2}) holds for arbitrary $\alpha$, 
comparing the coefficients of
different powers of $\alpha$ in the above Taylor series results 
in an {\it infinite} series of {\it exact } equations. 
Comparison of the coefficients of $\alpha ^0$ gives 
\be
{d\ln \langle S \rangle _g \over dg} = \langle R^d \rangle _g \qquad .
\label{gamma}
\ee
This is exactly the 
``gamma''-equation derived in \cite{ gamma}. This equation 
is valid not only for the BS model but for the Sneppen model and Invasion 
Percolation as well \cite{ PMB} since
it does not rely on the assumption that the sizes of subsequent avalanches
are uncorrelated.
To show this, one has to take a large number $N$ of $f_o$-avalanches and
write the balance equation of how this number decreases as $f_o$ is 
increased \cite{ PMB}.
Changing the variables from $g$ back to $f_o$ gives the more familiar
form of the ``gamma''-equation:  
${d\ln \langle S \rangle _{f_o} / df_o} = {\langle R^d \rangle 
 _{f_o} /(1-f_o)}$. 

Higher powers of $\alpha$ in Eq. (\ref{e.m2}) give
new {\it exact} equations.
Here we show just the first two:
\begin{eqnarray}
 {d \over dg} 
\left( {\langle S^2 \rangle _g \over \langle S \rangle _g} \right)
= 2\langle R^d S \rangle _g  \qquad ; \label{e.a1}\\
 {d \over dg} 
\left( {\langle S^3 \rangle _g \over 3 \langle S \rangle _g} 
- {\langle S^2 \rangle _g ^2 \over 2 \langle S \rangle _g ^2}\right)
= \langle R^d S^2 \rangle _g  \qquad . \label{e.a2}
\end{eqnarray}

The Taylor expansion changes slightly in the overcritical region, where 
there is a finite probability 
$P_{\infty}(g)$ to start an infinite avalanche. Since the avalanche
distribution $P(S, g)$ is limited to finite avalanches, 
it is naturally normalized to $1-P_{\infty}(g)$. Therefore,
when $g>g_c$ the Fourier series for $p(\alpha ,g)$ can be written as
$p(\alpha ,g)=1-P_{\infty}(g)-\langle S \rangle _g \alpha  + 
\langle S^2\rangle _g \alpha  ^2 /2 + \ldots $. Now the comparison 
of the coefficients at $\alpha ^0 $ in Eq. (\ref{e.m2}) gives
\be
{d \ln P_{\infty}(g) \over dg}=\langle R^d \rangle _g \qquad .
\label{e.beta}
\ee
This new equation is the $g>g_c$
analog of the ``gamma''-equation (\ref{gamma}). We will refer to it
as ``beta''-equation (the exponent $\beta$ is traditionally
used for the scaling of $P_{\infty}(g)$, while $-\gamma$ is used for
$\langle S \rangle _g$.) 

There is a more straightforward way to derive equation (\ref{e.beta})  
from the average properties of the merging process.
The merging of finite and 
infinite avalanches gives an infinite avalanche and, therefore, 
leads to an increase in $P_{\infty}(g)$. A simple analysis
gives an equation governing this process
as $d P_{\infty}/dg=\langle R^d 
\rangle _g P_{\infty}$, which is just Eq. (\ref{e.beta}).

As in the undercritical case, the
Taylor expansion of Eq.(\ref{e.m2}) for $g>g_c$ 
determines an infinite series of exact equations.
Here are the first two:
\begin{eqnarray}
{d \over dg}\left( {\langle S \rangle _g \over P_{\infty}(g)}
\right) = -\langle 
R^d S \rangle _g \qquad ;\\
{d \over dg}\left( {\langle S^2 \rangle _g \over P_{\infty}(g)}
+{2\langle S \rangle _g ^2 \over P_{\infty}(g) ^2}
\right) = -\langle 
R^d S^2 \rangle _g \qquad \label{e.a4}.
\end{eqnarray}

As in other creation-annihilation branching 
processes, the avalanche distribution $P(S, g)$ 
in BS model for $g<g_c$ is known to have a scaling form 
\be 
P(S,g)=S^{-\tau} f(S^{\sigma}(g-g_c)) \qquad , 
\label{scaling}
\ee
where $\tau$ and $\sigma$ are some critical exponents and $f(x)$ 
is a scaling function that rapidly decays to 
zero as $x \rightarrow -\infty$.  From (\ref{scaling}) 
it follows that the average avalanche size diverges when $g$
approaches $g_c$ from below as $\langle S \rangle _g \sim
(g_c-g)^{-\gamma}$, where $\gamma={2-\tau \over \sigma}$.
Substitution of this expression into the ``gamma''-equation
(\ref{gamma}) results in 
\be
\langle R^d \rangle _g = {\gamma \over g_c-g}, \qquad {\rm for} \quad g<g_c.
\label{e.gamma2}
\ee
In the BS model the critical exponent $\gamma$ determines
the amplitude of $\langle R^d \rangle _g$ below $g_c$.
The exponent relation derived from (\ref{e.gamma2}) 
connects $\tau$ to $D$ and $\sigma$: 
$\tau=1+d/D-\sigma$ \cite{ PMB}.
It is easy to see that Eqs (\ref{e.a1}-\ref{e.a4})
do not yield additional exponent relations 
but further restrict the exact form of the avalanche distribution.
In fact it can be shown that Eq. (\ref{e.m2}) 
and the exponents $D$ and $\tau$ uniquely determine
the shape of the scaling function $f(x)$ \cite{ tbp}.

The scaling should work in the overcritical regime as well.
However, unlike in the equilibrium statistical mechanics, 
the critical exponent $\sigma$ can a priori be different
above and below the transition. 
In what follows we show that at least for 
the BS model this is not true.
Substitution of the scaling form
form $P_{\infty}(g) \sim (g-g_c)^{\beta}$ into the ``beta''- equation 
(\ref{e.beta})
results in 
\be
\langle R^d \rangle _g = {\beta \over g-g_c}, \qquad {\rm for} \quad g>g_c.
\label{e.beta2}
\ee
Again, similar to the ``gamma''-equation (\ref{e.gamma2}),  
the critical exponent $\beta$ gives 
the amplitude of $\langle R^d \rangle _g$
above the transition. 
From (\ref{e.beta2}) it follows that the same 
exponent relation $\tau=1+d/D-\sigma$ holds in the overcritical
region, and, therefore the exponent $\sigma$ is the same above and
below the transition.
The scaling form (\ref{scaling}) 
can now be extended to include the overcritical region. For this one 
just lets the argument $x$ of the scaling function $f(x)$ be 
positive. 
As in various percolation problems \cite{ stauffer} the
scaling form (\ref{scaling}) for 
$P(S, g)$ at $g>g_c$ 
results in the exponent relation $\beta={\tau-1 \over 
\sigma}$.  An interesting consequence of exact Eqs.  (\ref{e.gamma2}, 
\ref{e.beta2}) is that the universal amplitude ratio for $\langle 
R^d \rangle _g$ is given by the ratio of two critical exponents
\be 
{\langle R^d \rangle _{g+\Delta g} \over 
\langle R^d \rangle _{g-\Delta g} }={\beta \over \gamma}=
{\tau-1 \over 2-\tau}
\ee
This unusual relation between the universal amplitude ratio and critical 
exponents is to our knowledge unique for the BS model.

There is a case when the master equation (\ref{e.m2}) can 
be written in a closed form. This is the
extensively studied \cite{ mf, sb} 
mean field random neighbor version of the BS model, where 
at each time step $K-1$ ``neighbors'' of the active site 
are selected in an annealed random fashion 
throughout the whole system. It is easy to see that in the thermodynamic 
limit of this model the number of updated sites
in the avalanche of temporal duration $S$ is given by 
$n_{cov}=(K-1)S+1$. 
This is the quantity that should be used instead of 
$R^d(S,g)$ in our equations. The missing equation connecting
$r(\alpha ,g)$ and $p(\alpha ,g)$ is $r(\alpha ,g)=-(K-1) {\partial p(\alpha ,g) \over 
\partial \alpha  }+ p(\alpha ,g)$ and the final form of the Eq. 
(\ref{e.m2}) for the mean field BS model is
\be
{\partial \ln(1-p(\alpha ,g)) \over \partial g}=-(K-1)
{\partial p(\alpha ,g) \over \partial \alpha } +p(\alpha ,g)
\label{e.m.mf}
\ee
This equation should be solved with the initial condition  
$p(\alpha ,0)=e^{-\alpha}$, since  $P(S,0)=\delta_{S,1}$.
We checked that for $K=2$ the generating function 
$\sum_{S=1}^{\infty} P(S,f_o)x^S= 
{1-2xf_o(1-f_o)-[1-4xf_o(1-f_o)]^{1/2} \over 2f_o^2x}$, 
derived in \cite{ sb} using different methods, 
after the substitution of $x=e^{-\alpha}$ and $f_o=1-e^{-g}$
satisfies (\ref{e.m.mf}) and has the correct initial condition.
That confirms the overall consistency of our approach.

\begin{figure}
\epsfxsize 8cm
\centerline{\epsfbox{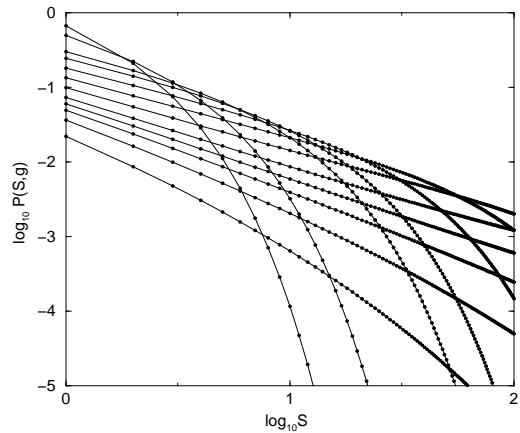}}
\caption{
The results of the numerical solution of Eq.(1)
on the interval $1 \leq S \leq 100$
with $R^d(S, g)=S^{0.412}$. Values of $g$ increase from 
top to bottom. The exponent of power law part was measured to be
$1.1 \pm 0.1$.
}
\label{one}
\end{figure}

It can be shown \cite{ tbp}
that the substitution of the scaling form (\ref{scaling})
into the Eq.(\ref{e.m2}) defines recursively all terms in
the Taylor series of the scaling function $f(x)$ at $x=0$:
%\be
$f^{(n+1)}(0)=\sum_{n_1+n_2=n}
{\Gamma (1-\tau +\sigma n_1) \Gamma (\sigma +\sigma n_2) \over
\Gamma (1-\tau +\sigma +\sigma n)}{n! \over n_1! n_2!}
f^{(n_1)}(0) f^{(n_2)}(0)$
%\ee
, where $\Gamma (x)$ is the Euler's gamma function.
 Actually Eq. (\ref{e.m2}) does more than that:
for a given $d/D$ it uniquely selects
$\tau$. 
We suspect that only for this $\tau$ the scaling 
function satisfies all usual requirements, such as $f(x) \rightarrow 0$, 
when  $x \rightarrow \pm \infty$, and $\int_0^{\infty} x^{-\tau}(
f(0)-f(-x^{1/\nu}))dx=0$ \ (the absence of the infinite avalanche below
$g_c$.) Which of these constraints actually defines 
$\tau$ as a function of $d/D$ remains to be determined.
The numerical solution of the Eq. (\ref{e.m1}) with
$R^d(S, f_o)=AS^{d/D}$ indeed seems to give the correct value
for $\tau$ \cite{ tbp}. In Fig. \ref{one} we present the results
of the numerical solution of Eq.(\ref{e.m1}) with $R^d(S, f_o)=
S^{0.412}$, corresponding to the best numerical estimate
of $1/D$ in the one-dimensional BS model
\cite{ grass, PMB}. The solution indeed yields
$\tau=1.1 \pm 0.1$ which is consistent
with $\tau = 1.07 \pm .01$ determined by
extensive Monte-Carlo simulations.

In \cite{ PMB} it was shown that all the exponents of a general 
extremal model
are determined in terms of just two independent ones, say $D$ and 
$\tau$.  As a result of the approach described here
the set of independent exponents for the BS model
was narrowed down to just one. It is tempting
to extend these arguments to other extremal models.
The weak point for this lies in the assumption that sizes of two
subsequent avalanches are mutually uncorrelated. 
At present it is unclear how strong is this correlation 
and how it influences the scaling.

This work was supported by the U.S. Department of Energy Division
of Material Science, under contract DE-AC02-76CH00016.
The author thanks Prof. Y.-C. Zhang and the 
Institut de Phisique Th\'{e}orique, Universit\'{e} de Fribourg
for hospitality during the visit when part of this work was accomplished.
The author is grateful to P. Bak and M. Paczuski for 
helpful comments on the manuscript.

\end{document}